\documentclass[12pt]{iopart}

\usepackage{graphicx}
\usepackage{dcolumn}
\usepackage{bm}

\newcommand{\glue}{$\tilde{\Pi}(\omega)$ }
\newcommand{\Sig}{$\Sigma(\omega)$ }
\newcommand{\biduo}{Bi$_{2}$Sr$_{2}$CaCu$_{2}$O$_{8+\delta}$ }
\newcommand{\otau}{$1/\tau(\omega)$ }
\newcommand{\mstar}{$m^{*}(\omega)/m_{b}$ }
\newcommand{\sig}{$\sigma_{1}(\omega)$ }
\newcommand{\sigopt}{$\hat{M}(\omega)$ }

\begin{document}
\title[Doping dependent optical properties of Bi2201.]{Doping dependent optical properties of Bi2201.}
\author{E. van Heumen$^1$, W. Meevasana$^3$, A.B. Kuzmenko$^1$, H. Eisaki$^2$, D. van der Marel$^1$}
\address{$^1$ D\'epartement de Physique de la Mati\`ere
Condens\'ee, Universit\'e de Gen\`eve, quai Ernest-Ansermet 24,
CH1211 , Gen\`eve 4, Switzerland}
\address{$^2$ Nanoelectronics Research Institute, National
Institute of Advanced Industrial Science and Technology,
Tsukuba, Japan}
\address{$^3$ Department of Physics, Applied Physics, and
Stanford Synchrotron Radiation Laboratory, Stanford University,
Stanford, CA 94305}
\ead{dirk.vandermarel@unige.ch}
\pacs{71.10.Ay, 72.80.Ga, 74.25.Gz, 74.72.Hs}
\submitto{\NJP}

\begin{abstract}
An experimental study of the in-plane optical conductivity of
(Pb$_{x}$,Bi$_{2-x}$)(La$_{y}$Sr$_{2-y}$)CuO$_{6+\delta}$
(Bi2201) is presented for a broad doping and temperature range.
The in-plane conductivity is analyzed within a strong coupling
formalism. We address the interrelationship between the optical conductivity ($\sigma(\omega)$), the single particle self energy, and the electron-boson spectral function. We find that the frequency and temperature
dependence can be well described within this formalism. We
present a universal description of optical, ARPES and tunneling
spectra. The full frequency and temperature dependence of the
optical spectra and single particle self-energy is shown to
result from an electron-boson spectral function, which shows a
strong doping dependence and weak temperature dependence.
\end{abstract}

\maketitle

\section{Introduction}
Many properties distinguish high temperature superconductors
from conventional superconductors, most importantly of course
the critical temperature. It is therefore not surprising that
experimental and theoretical efforts are aimed at finding
alternatives for the conventional electron-phonon coupling
driven pairing mechanism. The field is currently divided into
two schools. The first school explains the peculiarities of
these materials by strong correlation effects
\cite{anderson-science-1987,seibold-PRL-2005,phillips-annphys-2006,haule-PRB-2007,eichenberger-PRB-2008}
while the second seeks the origin in the coupling of electrons
to a spectrum of bosons
\cite{varma-PRL-1989,millis-PRB-1990,demler-nat-1998,hwang-nat-2004,norman-PRB-2006}.
The first school typically approaches the problem starting from
doping the Mott-insulating state of the parent compounds. The
resulting state of matter is not adiabatically connected to
another well known state of matter, the Fermi liquid, hence its
properties are fundamentally different. The important
interactions are then the on-site Coulomb repulsion $U$ and the
anti-ferromagnetic exchange coupling $J$, neither of which
qualifies as a retarded interaction as in the standard
Eliashberg framework \cite{anderson-SC-2007}. The second school
starts by assuming that the underlying state of matter is a
Fermi liquid but that the peculiar properties arise due to a
strong coupling of electrons to bosons different from phonons
in ordinary metals, for example spin fluctuations.

A problem in the study of the electron-boson coupling in
cuprates is the high critical temperature of these materials:
since spectroscopic signatures of the interaction are smeared
out by thermal broadening they are more easily determined at
low temperatures. However, a careful description of the
superconducting (d-wave) state complicates the analysis
tremendously. Nevertheless, some attempts have been made to
extract the bosonic spectral function in the superconducting
state \cite{hwang-PRB-2007}. The study of cuprates with a low
critical temperature can therefore be of interest, because the
analysis can be performed at low temperature without the
intervention of the superconducting state.

In a recent study we reported on an analysis of the optical
properties of several cuprates \cite{heumen-sub-2008}. We found
evidence for a broad, structured bosonic spectrum extending up
to energies of 300 to 400 meV. The spectra consist of two
features: (i) a robust peak in the range of 50 to 60 meV and
(ii) a doping dependent continuum extending to 300 - 400 meV
for the samples with the highest $T_c$. Interestingly, the
doping dependence of these spectra correlates with that of the
critical temperature. In this paper we report on the optical
spectra of
(Pb$_{x}$,Bi$_{2-x}$)(La$_{y}$Sr$_{2-y}$)CuO$_{6+\delta}$
(Bi2201)used in that study and discuss the analysis of
\cite{heumen-sub-2008} in more detail.

The structure of the paper is as follows. In section \ref{Exp}
we discuss the experimental techniques and optical
conductivity. In section \ref{analysis} the strong coupling
formalism is introduced and we present the analysis of optical
spectra within this formalism. In section \ref{discussion} we
make a comparison with angle resolved photoemission
spectroscopy (ARPES) and tunneling experiments within the
context of a model unifying the results from both experimental
techniques.

\section{Experiments}\label{Exp}
The optical properties of four samples of $Pb$ and $La$
substituted Bi2201 are studied: underdoped
Bi$_{2}$Sr$_{1.25}$La$_{0.75}$CuO$_{6+\delta}$ (UD0, $T_{c}=$ 0
K) and
Pb$_{0.5}$Bi$_{1.55}$Sr$_{1.2}$La$_{0.8}$CuO$_{6+\delta}$
(UD10, $T_{c}=$ 10 K), optimally doped
Pb$_{0.55}$Bi$_{1.5}$Sr$_{1.6}$La$_{0.4}$CuO$_{6+\delta}$
(OpD35, $T_{c}=$ 35 K) and overdoped
Pb$_{0.38}$Bi$_{1.74}$Sr$_{1.88}$CuO$_{6+\delta}$ (OD0,
$T_{c}=$ 0 K). Temperature dependent experiments are performed
in cryostats that have been specially adapted to keep the
sample position fixed during thermal cycling. Far-infrared
reflectivity experiments are carried out at pressures of about
$p\approx10^{-6}$ mbar while the mid-infrared reflectivity and
near infrared/ ultra violet ellipsometry experiments are
carried out under ultra high vacuum conditions with
$p\approx10^{-9}$ mbar. Samples are freshly cleaved just before
being inserted into the cryostat. For each sample, spectra were
taken between 10 K and 300 K with a temperature interval of 2
K. We use a Fourier transform infrared spectrometer for
reflection experiments in the range 5 meV to 750 meV. The
reference for the reflection measurements is taken on a gold
film deposited \textit{in-situ} on the sample. Ellipsometric
measurements are performed in the energy range 0.75 - 6 eV.

\subsection{Normal incidence reflectivity.}
\begin{figure}[!t]
\centering\includegraphics[width=8.5 cm]{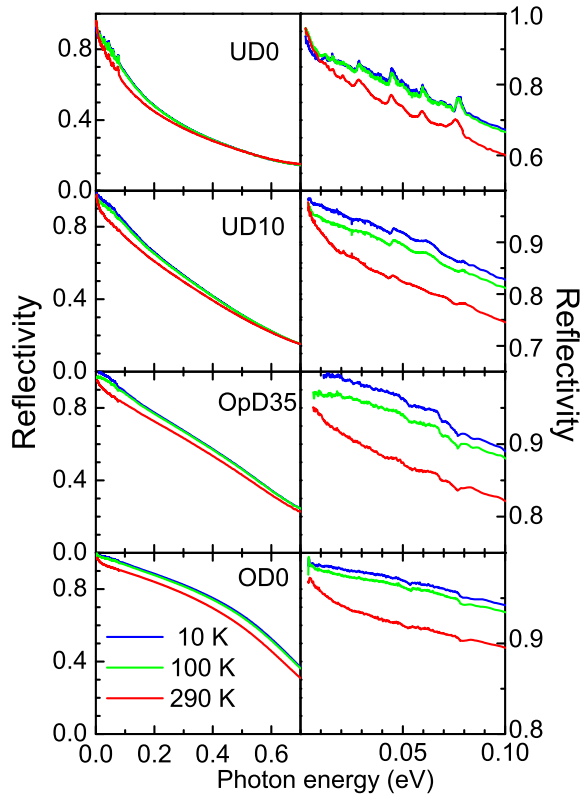}
\caption{\label{reflectivity} Reflectivity for 4 different dopings of Bi2201 at T = 10, 100 and 290 K.
>From top to bottom UD0, UD10, OpD35 and OD0 samples.
The left hand panels show the reflectivity on a large scale while the right hand panels show the low energy reflectivity highlighting the phonon range.}
\end{figure}
Reflectivity spectra are shown in figure \ref{reflectivity} for
all four samples at three temperatures: 10 K, 100 K and 290 K.
The reflectivity data is characteristic of a bad metal and
shows a strong doping dependence. With increasing doping the
metallicity increases and the reflectivity increases
accordingly. As a result the sharp phonon structures, which can
be clearly seen in the UD0 sample (top right panel) become
progressively more screened and are hardly visible in the OD0
spectra. The metallicity manifests itself in a dependence on
frequency that follows the Hagen-Rubens trend
$R(\omega)=1-2\sqrt{2\rho_{dc}\omega/\pi}$ for frequencies
below 25 meV (200 cm$^{-1}$). We made fits to the low frequency
reflectivity using the Hagen-Rubens expression to estimate the
d.c. resistivity. The values obtained in this way are in
agreement with those reported in the literature for similar
doping values \cite{ando-PRB-2000,ono-PRB-2003} and the fits
extrapolate to $R(\omega\rightarrow0)$ = 1 within our
experimental uncertainty ($\pm$ 0.5 \%). The exception is the
UD0 sample. On a qualitative level the reflectivity looks
similar to that of a disordered metal (for an example see
\cite{heeger-PRB-1993}). The temperature dependence is also
somewhat different from the other samples: the reflectivity
increases with decreasing temperature but below 10 meV the
reflectivity first shows a small increase and then starts to
decrease for temperatures below about 100 K.

\subsection{Ellipsometry}
\begin{figure}[!t]
\centering\includegraphics[width=8.5 cm]{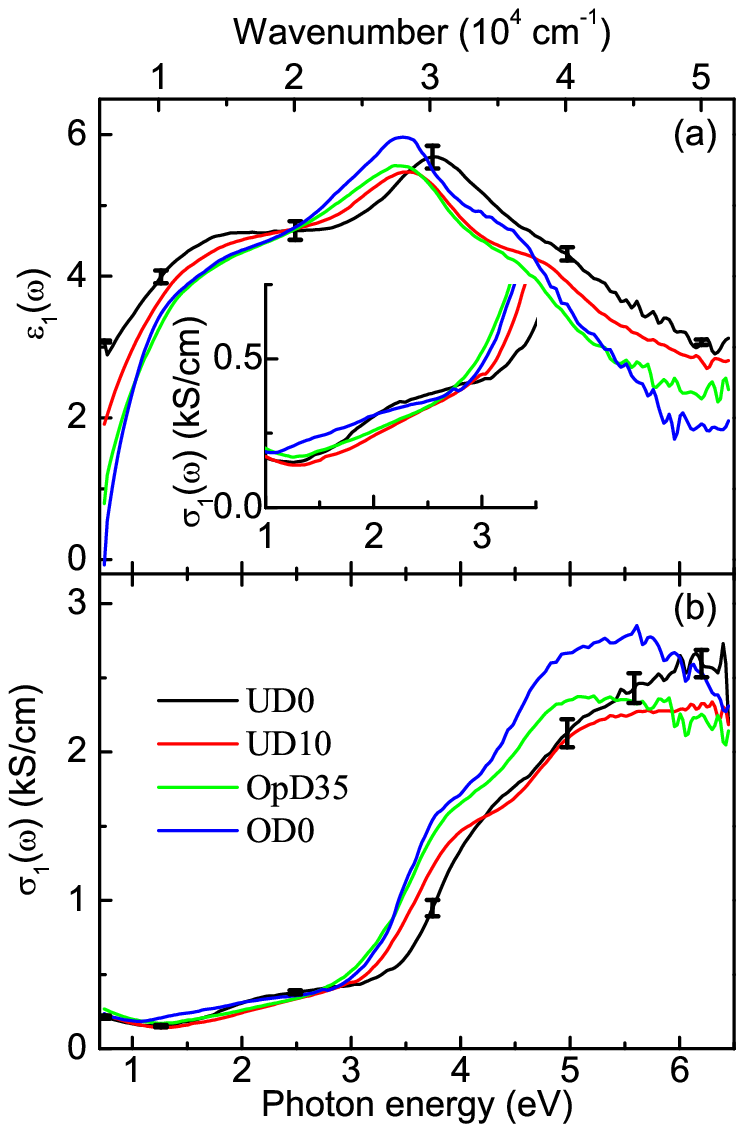}
\caption{\label{ellipso} (a): Real part of the dielectric function, $\varepsilon_{1}(\omega)$.
(b) Real part of the optical conductivity, $\sigma_{1}(\omega)$.
We only display room temperature spectra, since the temperature dependence for these frequencies is almost indiscernible at this scale.
The error bars shown for the UD sample are systematic error bars and are representable for all four samples.}
\end{figure}
Ellipsometric measurements give direct access to the complex
pseudo-dielectric function,
$\hat{\varepsilon}_{pseudo}(\omega)$, that corresponds closely
to $\hat{\varepsilon}_{ab}(\omega)$ with a weak admixture of
the $c$-axis component. Correspondingly, the transformation
from $\hat{\varepsilon}_{pseudo}(\omega)$ to
$\hat{\varepsilon}_{ab}(\omega)$ using the Fresnel equations
gives rise to a small correction for the $c$-axis admixture in
this frequency range. This has been verified for several high
T$_{c}$ cuprates with roughly the same optical constants
\cite{carbone-PRB-2006b,heumen-PRB-2007}. The $c$-axis
dielectric constant is almost independent of energy between
1.25 eV and 3 eV and equals roughly 4.1 for Bi2223 and 4.0 for
Hg-1201. We use a frequency independent $c$-axis dielectric
constant for Bi2201, $\hat{\varepsilon}_{c}\approx$ 4.3, to
correct $\hat{\varepsilon}_{pseudo}(\omega)$. A systematic
error for this procedure was determined by varying the value of
$\hat{\varepsilon}_{c}$ between 4 and 4.5 and is indicated as
the error bar in figure \ref{ellipso}.

The $c$-axis corrected functions obtained from the
ellipsometric measurements, $\varepsilon_{1,ab}(\omega)$ and
$\sigma_{1,ab}(\omega)$, are displayed in figure \ref{ellipso}.
We observe a series of interband transitions starting around 2
eV. At this energy there is a weak onset, which is most clearly
pronounced for the UD0 sample, which has been associated with
the onset of charge transfer excitations
\cite{uchida-PRB-1991}. We observe three more transitions at
higher energy which show a significant doping dependence.

\subsection{In-plane conductivity}
\begin{figure*}[htb]
\centering
\includegraphics[width=15 cm]{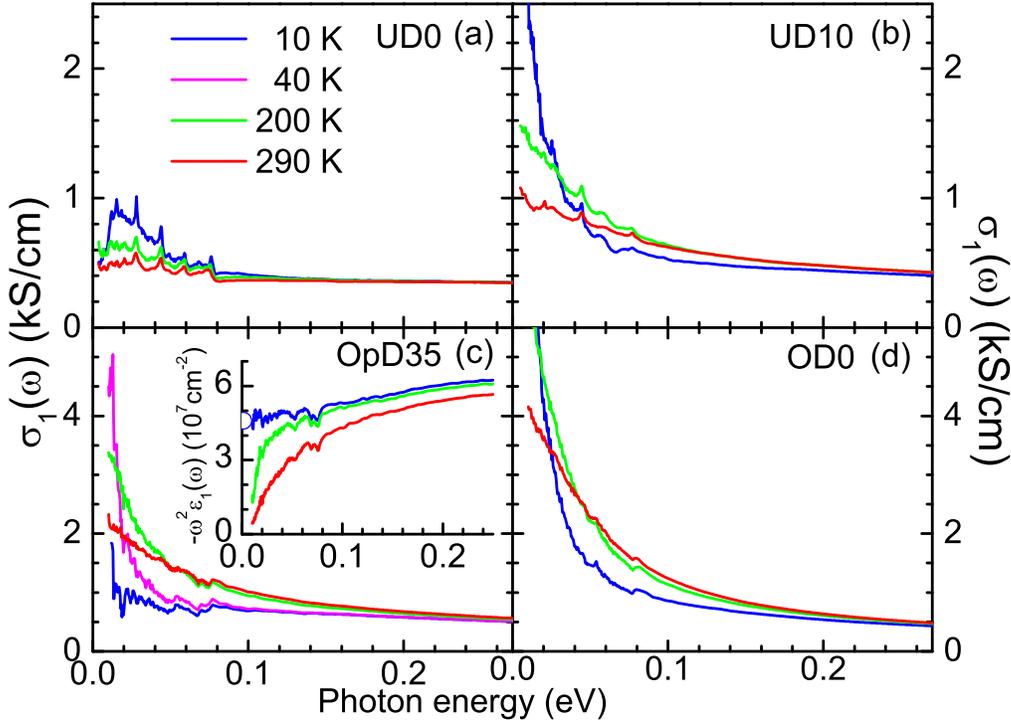}
\caption{\label{sigma} Optical conductivities for (a): UD0,
(b): UD10, (c): OpD35 and (d): OD0 samples. Note the difference
in scale between the upper and lower panels. For the OpD35
sample an extra curve is shown just above T$_{c}$. The inset in
panel (c) shows an estimate of the superfluid density for the
OpD35 sample. The open semi circle corresponds to the estimate
found from a Drude-Lorentz fit of the data.}
\end{figure*}
The combination of reflection experiments and ellipsometry
makes the determination of the optical conductivity in the IR
range more robust, as detailed in \cite{kuzmenko-RSI-2005}. We
display the optical conductivity, $\sigma_{1}(\omega)$, in
figure \ref{sigma} at several temperatures. As expected the low
frequency conductivity increases strongly with doping (note the
factor of two difference in scale between the upper and lower
panels). With the exception of the UD0 sample, the
conductivities show a metallic Drude response. The UD0 sample
shows a broad incoherent response with an interesting
temperature dependence: with decreasing temperature an
asymmetric peak develops with a maximum at 15 - 20 meV. A
similar structure and temperature trend has also been observed
for YBa$_{2}$Cu$_{3}$O$_{6+\delta}$ (YBCO)
\cite{lee-PRB-2005a}, indicating that this is a generic feature
of underdoped cuprates. The pronounced mid-infrared structure
observed in \cite{lee-PRB-2005a} is less visible here. The
optical conductivity of the OpD35 sample at T = 10 K does not
show a clear signature of the superconducting gap, but the
conductivity is strongly suppressed as compared to \sig at T =
40 K. Another signature of superconductivity is the presence of
a superfluid density, which appears as a
$\delta(\omega)$-function in $\sigma_{1}(\omega)$ with strength
$\omega_{p,s}$. $\omega^{2}_{p,s}$ is directly proportional to
the superfluid density and can be estimated from a
Drude-Lorentz fit. We find $\omega_{p,s}\approx$ 0.84 eV. As a
double check we use the fact that, due to causality, the delta
function in $\sigma_{1}(\omega)$ gives rise to a contribution
to
$\varepsilon_{1}(\omega)\approx-\omega_{p,s}^{2}/\omega^{2}$.
The inset in figure \ref{sigma}c compares the superfluid
density obtained by extrapolating
$-\omega^{2}\varepsilon_{1}(\omega)$ to zero frequency to the
one obtained from the Drude-Lorentz fit, shown as a semi
circle. The two estimates are in good agreement. The superfluid
density is substantially smaller compared to optimally doped
Bi2212 \cite{molegraaf-science-2002} ($\omega_{p,s}\approx$
1.18 eV) and Bi2223 \cite{carbone-PRB-2006b}
($\omega_{p,s}\approx$ 1.28) but  T$_{c}$ is correspondingly
lower. This correlation between T$_{c}$ and the superfluid
density has been observed by several authors
\cite{uemura-PRL-1989,homes-nat-2004}.

\section{Analysis}\label{analysis}

\subsection{Extended Drude model}
The Drude model describes the optical conductivity of a gas of
non-interacting electrons with a single, energy independent
channel of dissipation:
$4\pi\hat{\sigma}(\omega)=\omega_{p}^{2}/(1-i\omega\tau)$. This
model is insufficient to describe the situation where the
dissipation in the electron system arises from the
electron-phonon or electron-electron interactions. To deal with
these situations we can generalize the Drude model by
introducing a memory function $\hat{M}(\omega)$
\cite{allen-PRB-1977},
\begin{equation}\label{gendrude}
\hat{\sigma}(\omega)\equiv\frac{\omega_{p}^{2}}{4\pi}\frac{i}{(\omega+\hat{M}(\omega))},
\end{equation}
or ``optical self-energy''
($2\hat{\Sigma}_{op}(\omega)=-\hat{M}(\omega)$)
\cite{hwang-nat-2004}.Two related quantities are the effective
mass, $m^{*}(\omega)/m=M_{1}(\omega)/\omega+1$ and scattering
rate, $1/\tau(\omega)=M_{2}(\omega)$. These quantities are
useful because they are more easily interpreted physically.
(\ref{gendrude}) is easily inverted to obtain the scattering
rate \otau and effective mass \mstar directly from the measured
conductivity, provided one has an estimate for the plasma
frequency, $\omega_{p}$ and the contribution due to interband
transitions $\varepsilon_{\infty}$
\cite{dirk-nat-2003,heumen-PRB-2007}.
\begin{figure}[thb]
\centering\includegraphics[width=8.5 cm]{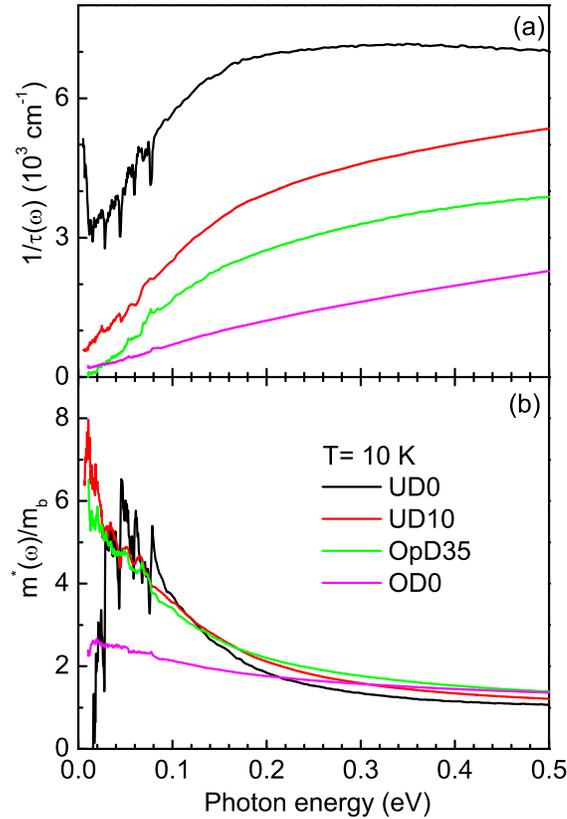}
\caption{\label{extdrude}  Scattering rate, \otau and effective mass, \mstar at T = 10 K.
Note the upturn in the UD0 scattering rate which indicates that the data cannot be described within this formalism.}
\end{figure}
\otau and \mstar are shown in figure \ref{extdrude} at T =10 K.
The UD0 sample has the highest scattering rate and strongest
frequency dependence. However, \otau shows an upturn at low
frequency and a corresponding negative effective mass in the
same frequency range. The implication of this is that the low
temperature, low frequency properties cannot be described in
terms of an interacting gas of electrons and thus invalidates
the extended Drude analysis for this compound. A similar
observation was made in \cite{lee-PRB-2005a} for heavily
underdoped YBCO samples. However the scattering rate reported
there shows a peak structure around 150 meV which is not found
here.

The other three samples show monotonically increasing
scattering rates, except in the vicinity of direct optical
phonon absorptions. The overall scattering rate as well as the
magnitude of the frequency dependence strongly decreases with
increasing doping. This can be related to a decreasing coupling
to low energy bosonic excitations. A measure of this coupling
strength can be obtained from the relation
$m^{*}/m_{b}(\omega\rightarrow0)=\lambda+1$. Therefore
$\lambda$ can be estimated directly from figure
\ref{extdrude}b, showing that $\lambda$ decreases from
$\lambda\approx$ 6.5 for UD10 to $\lambda\approx$ 1.5 for the
OD0 sample. The room temperature values for the three samples
are $\lambda_{UD10}\approx$ 4.5, $\lambda_{OpD35}\approx$ 3 and
$\lambda_{OD0}\approx$ 1.5. The determination of these coupling
strengths from experimental quantities will serve as an
important consistency check in section \ref{strong}.

\subsection{Strong coupling formalism}\label{strong}
In this section we will focus on the memory function because of
its relation to self-energies above and below the Fermi energy
and because structures arising from electron-boson interactions
are more clearly visible in $M_{1}(\omega)$ than in \mstar and
\otau \cite{hwang-nat-2004,norman-PRB-2006}. Encoded in the
memory function are the self-energy effects associated with the
electron-electron and electron-phonon interaction. The relation
between single particle self-energies and the memory function
is given by
\cite{pballen-PRB-1971,pballen-book-1982,mahan-book},
\begin{equation}\label{SEopt}
\frac{\hat{M}(\omega,T)}{\omega}=\left[\int\limits_{-\infty }^{+\infty } \frac{{[f(\omega  + \varepsilon,T) - f
(\varepsilon,T)]d\varepsilon}}{{\omega  - \Sigma (\varepsilon + \omega ,T) +
\Sigma ^* (\varepsilon,T)}} \right]^{-1}-1,
\end{equation}
where $f(\varepsilon,T)$ are Fermi factors and $\Sigma(\omega)$
and $\Sigma^{*}(\omega)$ are self-energies above and below the
Fermi energy. To arrive at this result we have to make some
assumptions. First, we neglect vertex corrections and the
energy dependence of the density of states. Furthermore the
self-energies are assumed to be momentum independent. We will
return to this latter assumption in section \ref{discussion}.
Under these circumstances we can express the single particle
self-energy as,
\begin{equation}\label{SEeq}
 \Sigma \left({\omega ,T}\right) =  \int d\varepsilon \int d\omega' \tilde{\Pi}(\omega')
 \left[\frac{n(\omega') + f(\varepsilon)}{\omega - \varepsilon + \omega' + i\delta}
  + \frac{n(\omega') + 1 - f(\varepsilon)}{\omega - \varepsilon - \omega' - i\delta }\right]
\end{equation}
with Bose factors, $n(\varepsilon,T)$ and the``glue function'',
$\tilde{\Pi}(\omega)$. The tilde on this quantity indicates
that it is an effective spectral density that potentially
contains features not arising from coupling to bosons, for
example the opening of a pseudogap in the density of states.
Several authors have discussed the applicability of this
formalism and its underlying assumptions to the cuprate problem
\cite{anderson-SC-2007,maksimov-review-2000,sharapov-PRB-2005,chubukov-PRB-2005,maier-prl-2008}.
We want to determine whether or not it can be consistently
applied to the optical properties of cuprates and see the
implications of the resulting glue-functions with regard to
other experimental probes.

Several microscopic theories give predictions about the shape
of \glue. We have tested these predictions, as well as
``model-independent'' functions, for HgBa$_{2}$CuO$_{4+\delta}$
as described in \cite{heumen-LT-2008}. A general function that
consists of a histogram representation of \glue using N blocks
with adjustable positions and heights,
\begin{equation}\label{HG}
\tilde{\Pi}(\omega)=f_{i}\quad\omega_{i-1}\le\omega\le\omega_{i}
\end{equation}
where $i$ runs from 1 to N, $\omega_{0}$ = 0 and $f_{i}$ is the
height of the $i$th block. For
$\omega_{0}\le\omega\le\omega_{1}$ we use
$\tilde{\Pi}(\omega)=f_{1}\omega$ to circumvent the divergence
of the integral in (\ref{SEeq}). The 2$N$ parameters of this
model are optimized using a standard Levenberg-Marquardt least
square routine. This routine optimizes the parameters of the
histogram model to obtain the best fit to the experimental
reflectivity and ellipsometry data. The resulting histogram
gives an indication of the important features in the spectrum.
We find that the detail of \glue that can be extracted from the
experimental data is best represented with $N$ = 6 blocks. For
cuprates the function \glue consists of two main features: a
peak with an energy in the range 50 - 60 meV and a broad
spectrum extending up to 400 meV \cite{heumen-sub-2008}.

\subsection{Temperature dependence of the self-energy/memory
function} (\ref{SEopt}) and (\ref{SEeq}) depend on temperature
through the Bose and Fermi factors only. In the electron-phonon
problem the temperature dependence of the phonon spectral
density is very weak, and can be neglected. However, if the
bosons that make up \glue derive from electronic degrees of
freedom \glue can depend on temperature due to feedback effects
\cite{chubukov-PRB-2005}. In \cite{heumen-sub-2008} we found
that a fit of the experimental data at all temperatures
required a temperature dependent $\tilde{\Pi}(\omega)$,
suggesting that it derives at least in part from electronic
degrees of freedom.
\begin{figure}[tbh]
\centering\includegraphics[width =15 cm]{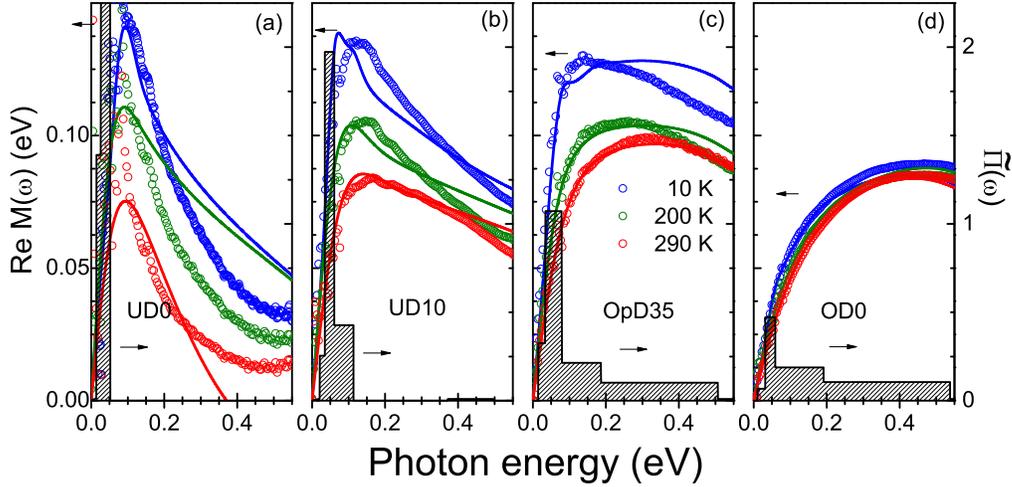}
\caption{\label{SE}  Memory function \sigopt (symbols) together with calculated quantities (solid lines).
The shaded area corresponds to the room temperature \glue function used in the calculations.}
\end{figure}
In figure \ref{SE} the room temperature \glue spectra are
reproduced together with experimental and calculated memory
functions for the four samples. The experimental memory
functions show a strong doping dependence from narrow and
strongly peaked in the underdoped regime to broad with no clear
maximum on the overdoped sample. The comparison with the \glue
function shows that the presence of this maximum correlates
with the strength of the peak at 50 - 60 meV in the \glue
function. We use the \glue spectrum that gives the best fit to
the room temperature optical spectra to calculate the memory
function at several temperatures. At room temperature the
calculated \sigopt is therefore the best fit to the
experimental one based on (\ref{SEopt}) and (\ref{SEeq}). The
reason for choosing the room temperature \glue is that at lower
temperatures the spectra might contain features related to the
pseudogap. The quality of the fit becomes progressively worse
with decreasing doping. For the most underdoped sample, UD0,
the room temperature data can no longer be described by this
formalism. So we may conclude that (i) (\ref{SEopt}) and
(\ref{SEeq}) describe well the state of affairs in optimally
and overdoped superconductors and (ii) the equations fail
miserably for the most underdoped sample. This could imply a
doping induced transition between two different states of
matter, non-Fermi liquid on the underdoped side and Fermi
liquid on the overdoped side. The exact doping value separating
these two regimes is difficult to establish beyond any question
with this method: the fits and temperature dependence may still
be rather good if the system is only slightly away from the
Fermi liquid regime.

\subsection{Temperature dependence of \glue}
\begin{figure}[h]
\centering
\includegraphics[width=8.5 cm]{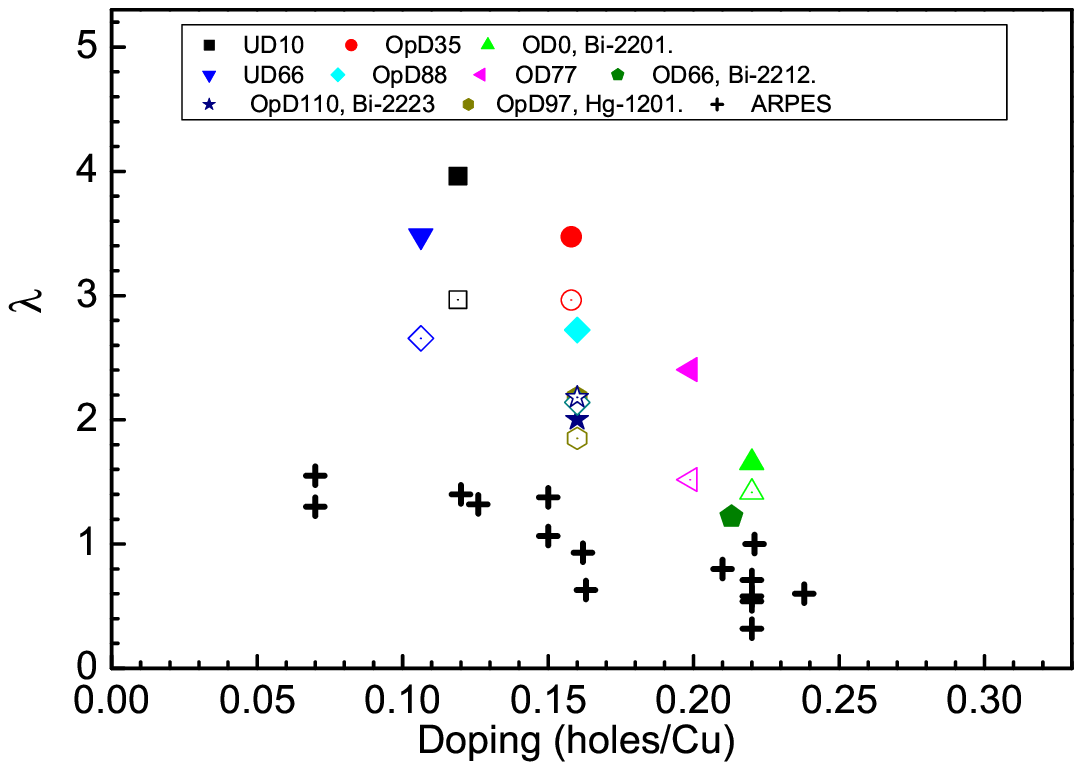}
\caption{Coupling strength as a function of carrier concentration.
Open symbols: 290 K, closed symbols: 100K. The crosses indicate the values obtained in \cite{lanzara-nat-2001}.
The observed trend of decreasing $\lambda$ with doping is consistent with the one obtained by ARPES
(see also \cite{hwang-PRB-2007,johnson-PRL-2001,non-PRL-2006})}
\label{lambda}
\end{figure}

\sigopt is calculated at 10 K and 200 K by correspondingly
changing temperature in (\ref{SEopt}) and (\ref{SEeq}) but
using the same \glue as for 290 K. It is clear from figure
\ref{SE} that most of the temperature dependence of the optical
spectra can indeed be explained within the strong coupling
formalism. The remaining differences (for example, at high
energies) can be removed by optimizing the parameters of the
histogram at each particular temperature.
\begin{figure}[tbh]
\centering\includegraphics[width =8.5 cm]{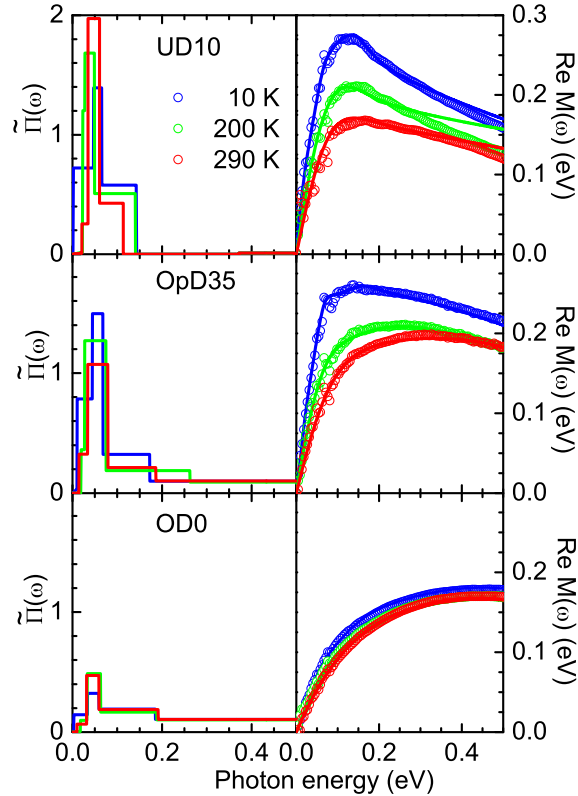}
\caption{\label{SE2}  Left panels: \glue functions obtained by optimizing the histogram parameters at each temperature.
Right panels: Corresponding \sigopt. Symbols are experiments and full lines calculations.}
\end{figure}
These results are shown in the panels of figure \ref{SE2} where
the memory functions are calculated from \glue functions at
corresponding temperatures. The temperature dependence of \glue
is weak for the OD0 sample but becomes stronger with decreasing
doping. The coupling constant, $\lambda$, can be obtained by
two methods (i) from $\lambda = m^{*}(\omega\rightarrow 0)/m-1$
as discussed above and (ii) using the relation
$\lambda=2\int_{0}^{\infty}\tilde{\Pi}(\omega)/\omega d\omega$.
We obtained the same values with these two methods. In figure
\ref{lambda} we show the coupling constants derived for the
Bi2201 spectra together with those derived from the other \glue
functions presented in \cite{heumen-sub-2008}.
The result shows a strong and systematic increase of $\lambda$
for decreasing hole concentration. This trend can also be
obtained from other probes, for example from ARPES as indicated
by the crosses which are taken from \cite{lanzara-nat-2001}.
The ARPES coupling constants are estimates for the nodal
direction. They represent only a partial coupling constant
determined from the change in dispersion around the low energy
kink and are therefore smaller than the ones determined from
optics. There is a trend of increasing coupling constant when
the temperature is reduced. This temperature dependence becomes
stronger upon decreasing the doping. The latter systematic
temperature dependence of $\lambda$ is a direct consequence of
a growth of intensity at low energies below the 50 - 60 meV
peak (see the \glue functions of the UD10 sample in the
left-hand panels of figure \ref{SE2}). As mentioned above, all
trivial thermal factors contained in (\ref{SEopt}) and
(\ref{SEeq}) are in principle folded out by our procedure. The
remaining temperature dependence reflects therefore the thermal
properties of the 'glue-function' itself. In a separate paper
\cite{heumen-sub-2008} we have shown for a large number of
cuprates, including the present set of samples, that reasonable
(albeit 2-3 times higher than experimental) values of T$_{c}$
are predicted by the Eliashberg equations using these \glue
functions. Moreover the 50 - 60 meV peak seems to be irrelevant
for the pairing in the overdoped samples.
In the Bi2201 materials T$_c$ is low, despite the fact that the
electron-boson coupling found in optical and ARPES experiments
is large. \glue is, in fact, quite similar to {\em e.g.} Bi2212
and Bi2223 where the maximum T$_c$ is of order 100 Kelvin. This
is a natural consequence of cation disorder at the Sr site in
the single layer Bi2201 compounds, suppressing rather
effectively the $d$-wave order parameter in the copper-oxygen
planes\cite{eisaki-PRB-2004}.

\subsection{Discussion.}\label{discussion}
Hwang \textit{et al.} have investigated the coupling to bosonic
modes by analyzing the optical scattering rates of Bi-2212
\cite{hwang-nat-2004,hwang-PRB-2007,hwang-PRB-2004,hwang-JPCM-2007,hwang-PRL-2007,hwang-PRL-2008b},
YBCO \cite{hwang-PRB-2006} and LSCO \cite{hwang-PRL-2008a}.
Several other groups have applied different methods of analysis
to optical spectra with similar results but differing
interpretations
\cite{thomas-PRL-1988,schlesinger-nat-1991,timusk-PRL-1991,basov-nat-1999,munzar-physc-1999,tu-PRB-2002,timusk-JPCM-2003,dordevic-PRB-2005,lee-PRB-2005b,casek-PRB-2005}.
Several other techniques, like ARPES
\cite{lanzara-nat-2001,johnson-PRL-2001,non-PRL-2006,bogdanov-PRL-2000,kaminski-PRL-2001,gromko-PRB-2003,borisenko-PRL-2003,zhou-nat-2003,cuk-PRL-2004,zhou-PRL-2005,valla-PRL-2007,inosov-PRL-2007}
and tunneling
\cite{zasad-PRL-1999,zasad-PRL-2001,lee-nat-2006,castro-PRL-2008},
also show evidence for coupling of electrons to bosonic modes.

\subsection{Comparison with other optical experiments.}
In most of the earlier work a spectral function consisting of
two contributions is reported: a peak below 100 meV and
structure in the 200 - 300 meV range. The doping trend of
\sigopt seen in figure \ref{SE} is similar to the one observed
in \cite{hwang-JPCM-2007} for \biduo, namely that \sigopt is a
broad and featureless function for high dopings and becomes
peaked and narrow for lower dopings. The authors of
\cite{hwang-PRB-2007} apply an analysis similar to ours and
also find an overall decrease in intensity in \glue with
increasing doping (note that in \cite{hwang-PRB-2007} the
notation $I^{2}\chi(\omega)$ is used for \glue). The main
difference with their analysis is that we find the spectra to
be much less temperature dependent in the low energy region of
\glue, in agreement with the results obtained by Dordevic
\textit{et al.} \cite{dordevic-PRB-2005}. Hwang \textit{et al.}
explain the temperature and doping dependence of these spectra
as arising from the presence of a pseudogap at lower
temperatures for the underdoped samples \cite{hwang-PRL-2008b}.
Although we cannot fully exclude this possibility, the analysis
in figure \ref{SE} shows that another interpretation is also
possible: namely the peak in \sigopt at low doping arises from
the coupling to a bosonic mode. This peak is smeared out at
higher temperatures by the Bose and Fermi factors appearing in
(\ref{SEopt}) and (\ref{SEeq}). The peak in \sigopt disappears
with increasing doping simply because the coupling to low
energy bosonic modes gets weaker. The temperature effect is
much more pronounced for Hg1201 where the peak in \sigopt is
more clearly visible \cite{heumen-sub-2008}. Although there is
no simple relation between the energy of a peak seen in \sigopt
and one seen in \glue a pseudogap would shift the peak in
\sigopt to higher energies as compared to the peak in \glue. We
note that the experimental \sigopt in figure \ref{SE} show the
opposite trend, i.e. the peak in \sigopt shifts to
\textit{lower} energy with decreasing temperature. Our analysis
thus shows that features in the optical spectra which have been
attributed to the opening of a pseudogap \cite{basov-RMP-2005}
can in fact be explained by a nearly temperature independent
peak in \glue in the 50 - 60 meV range.

\subsection{Comparison with ARPES.}
The experimental results from ARPES can be summarized as
indicative of two distinct structures in the electron energy
dispersion. Along the nodal direction one observes a kink at
low energy $\sim$ 50 - 70 meV
\cite{lanzara-nat-2001,non-PRL-2006,bogdanov-PRL-2000,kaminski-PRL-2001}
followed by a ``waterfall'' around 350 meV
\cite{valla-PRL-2007,inosov-PRL-2007}. The interpretation of
these two features is still under debate. Along the anti-nodal
direction only the low energy feature is observed since the
band bottom occurs around 100 meV. The kink has a smaller
energy $\approx$ 40 meV in the normal state and is shifted to
$\approx$ 70 meV in the superconducting state as a result of
the opening of the superconducting gap
\cite{gromko-PRB-2003,borisenko-PRL-2003,cuk-PRL-2004,zhou-PRL-2005}.

\begin{figure}[thb]
\centering\includegraphics[width=10 cm]{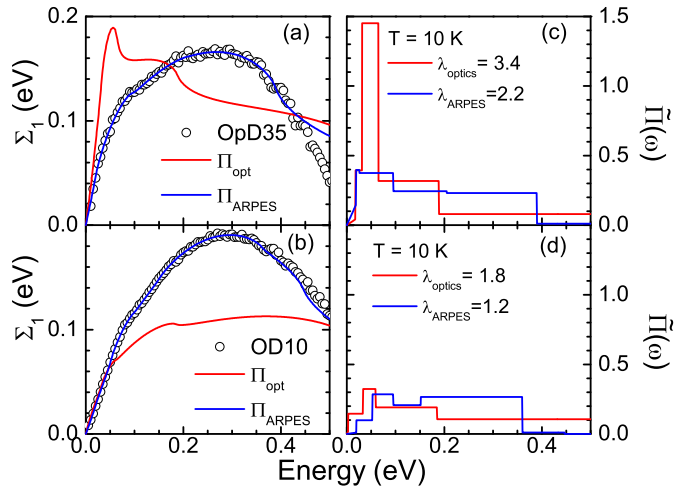}
\caption{\label{SEARPES} (a): Comparison between experimental \Sig (open circles),
calculated \Sig from the 10 K \glue spectra (red lines) and \Sig obtained from a direct fit to the ARPES data (blue lines) for the OpD35 sample.
(b): Same as in (a) but for the OD10 sample.
(c,d): The \glue functions corresponding to the calculated \Sig's in the left panels. Also indicated are the total coupling constants corresponding to these spectra.}
\end{figure}

If we ignore for the moment the momentum anisotropy, then the
self energies calculated with \glue can be compared to \Sig
extracted from ARPES. ARPES experiments have been performed on
crystals taken from the same batch as used in this study
\cite{non-PRL-2006,non-PRB-2008}. The self-energy along the
nodal direction has been determined using a full 2D analysis of
the spectral function using the LDA band structure as bare band
dispersion \cite{non-PRB-2008}. The self energies of the OpD35
and OD10 samples are shown in figure \ref{SEARPES}a and
\ref{SEARPES}b respectively by open circles. The self energies
calculated from the optical \glue function are shown in the
same figure in red. These self energies are quite different but
a closer examination reveals that several features correspond
quite well. Both experiments show structure in the 50 -60 meV
range and display a maximum in \Sig in the 200 - 300 meV range.
In order to determine the experimental \Sig from ARPES an
assumption has to be made for the bare band dispersion. The
fact that both the experimental and calculated self-energy have
almost parallel slope in the limit of frequency going to zero,
suggests that at least the low energy properties determined by
optical spectroscopy and ARPES are compatible.

\begin{figure}[thb]
\centering\includegraphics[width=8.5 cm]{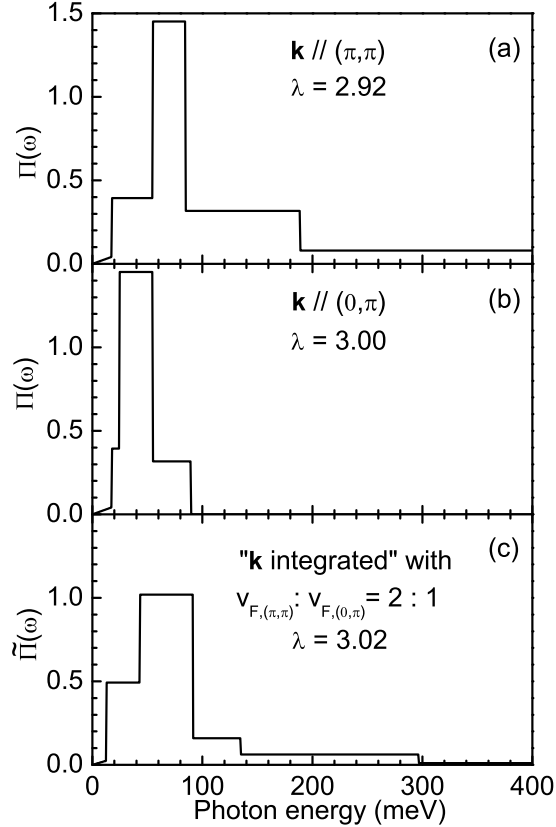}
\caption{\label{anisotropy} Estimate of the effect of k-space anisotropy on the optical conductivity. (a): Assumed \glue for the nodal direction.
The main peak is situated at 70 meV as reported in \cite{lanzara-nat-2001}. (b) Assumed \glue for the anti-nodal direction.
The main peak is situated at 40 meV as determined by ARPES \cite{gromko-PRB-2003}. Also indicated is the integrated coupling constant $\lambda$
for each spectrum.
(c) \glue determined by analyzing the optical conductivity assuming contributions to \sig from panels (a) and (b) weighted with a ratio of the nodal to antinodal Fermi velocity 2 : 1. }
\end{figure}

The ARPES self-energy can be further analyzed using our
histogram model for $\tilde{\Pi}(\omega)$. The fit to the ARPES
self-energy is shown in blue in figure \ref{SEARPES}a and
\ref{SEARPES}b. The \glue functions obtained from optics and
ARPES are compared in panels \ref{SEARPES}c and \ref{SEARPES}d.
The optical and ARPES glue-functions, although at first glance
quite different, have a number of globally similar features:
(i) the overall coupling strength, (ii) the broad energy range
and (iii) the 50 - 60 meV peak. We note that the low energy
kink in the ARPES data gives rise to a peak in \glue at the
same energy as the optical data. The total coupling constants,
indicated in figure \ref{SEARPES}c and \ref{SEARPES}d, are
somewhat larger for the optical \glue function. The same trend
is seen in figure \ref{lambda} where we compared $\lambda$
obtained from ARPES and optics for several samples. Since the
optical conductivity has contributions from all of k-space one
may expect that this explains the difference between optical
and ARPES results. Stojkovic and Pines show that the optical
conductivity is a linear superposition of contributions along
the Fermi surface where the self-energies are k-dependent
quantities \cite{stojkovic-PRB-1997},
\begin{eqnarray}
\hat{\sigma}(\omega)=\frac{e^{2}}{8\pi^{2}}\int_{FS}\frac{dk}{|v_{f}|}\int\limits_{-\infty }^{+\infty } \frac{d\varepsilon[f(\omega  + \varepsilon,T) - f
(\varepsilon,T)]}{i\omega}\times \nonumber \\
\frac{v_{f}^{2}}{\omega  - \Sigma_{k} (\varepsilon + \omega ,T) +
\Sigma_{k} ^* (\varepsilon,T)},
\end{eqnarray}
We can estimate the effect of anisotropy using our knowledge of
the electronic structure from ARPES. As summarized above, the
kink disperses from 70 meV at the nodal direction to 40 meV at
the anti-nodal direction \cite{zhou-review-2007}. The second
ingredient needed is the \textit{bare} Fermi velocity, $v_{f}$.
Estimates for the k-dependence of $v_{f}$ are available for
Bi-2212 from Kaminski \textit{et al.} \cite{kaminski-PRB-2005}.
The effect of anisotropy on \glue can now be estimated by
assuming that $\hat{\sigma}(\omega)$ has two contributions: one
from the nodal direction and one from the anti-nodal direction.
We separately calculate the conductivity arising from the nodal
and anti-nodal directions using the \glue functions given in
figure \ref{anisotropy}a and \ref{anisotropy}b. These functions
are chosen such that they are compatible with the experimental
observations discussed above. The resulting conductivities are
weighted with the Fermi velocity ratio $v_{f,(\pi,\pi)}$ :
$v_{f,(0,\pi)}$ = 2 : 1.
The resulting optical conductivity is then analyzed using the
same method as the experimental data in section \ref{strong}.
The \glue function shown in figure \ref{anisotropy}c shows that
the momentum anisotropy in \Sig gives rise to a mode energy
that is a weighted average of the nodal and antinodal
direction. A comparison of these results with the \glue
functions from figure \ref{SEARPES}c and \ref{SEARPES}d shows
that, (i) the low energy kink seen by optics has an energy
somewhat smaller than the nodal kink energy seen by ARPES and
(ii) there is less intensity in the high energy range seen by
optics compared with the nodal self-energy determined from
ARPES. Both effects arise because the optical \glue function is
an effective function that contains contributions from all
points around the Fermi surface.

\subsection{Comparison to tunneling experiments.}
The analysis of tunneling data for the cuprates turns out to be
more complicated than for conventional superconductors. Several
groups have found evidence for coupling to bosonic modes. Two
groups report coupling to a mode around 40 meV
\cite{zasad-PRL-1999,zasad-PRL-2001,castro-PRL-2008}. This is
significantly lower than our result. We note however that the
tunneling experiments are mainly sensitive to the anti-nodal
region of k-space due to the presence of a van Hove
singularity. As we have seen above ARPES experiments indicate
that in the antinodal region the mode energy is indeed close to
40 meV. However, using a different analysis Lee \textit{et al.}
report coupling to a bosonic mode at 53 meV in Bi2212
\cite{lee-nat-2006}.

\section{Conclusions and outlook.}
The optical properties of Bi2201 show an interesting evolution
with doping. The strongly underdoped UD0 sample shows a
non-Drude behavior at low temperatures, while the more doped
compounds all show Drude like behavior. It is interesting to
speculate at which doping the transition from one type of low
temperature behavior to the other occurs. One possibility is
that it coincides with the onset of superconductivity. An
interesting consequence of our analysis is than that the onset
of superconductivity separates two different normal states of
matter. The optical properties for higher dopings could be
described within the strong coupling formalism while the
properties of the UD0 sample clearly fall in a different
universality class dominated by strong electron correlations.

The strong coupling analysis shows that the coupling to bosonic
modes in these materials decreases strongly as a function of
doping. For the UD10 sample we find a narrow boson spectrum
with a large intensity in the range below 100 meV. With
increasing doping two separate trends are observed: the
spectrum broadens and the coupling to the 50 - 60 meV peak is
strongly reduced. Our study shows that the main temperature
dependence of the optical spectra can be understood to arise
from the thermal factors appearing in the strong coupling
formalism. To consistently fit the frequency dependence at all
temperatures we have to allow for temperature dependence in the
\glue function. This suggests that part of the \glue spectrum
is electronic in nature. Since the peak in the 50 - 60 meV
range is present at all temperatures and has a relatively weak
doping dependence, we conclude that it partially arises from
vibrational degrees of freedom.

\section{Acknowledgments}
We gratefully acknowledge stimulating discussions with J.
Zaanen, D.J. Scalapino, C.M. Varma, A.V. Chubukov, T.P.
Devereaux and Z.X. Shen. This work is supported by the Swiss
National Science Foundation through Grant No. 200020-113293 and
the National Center of Competence in Research (NCCR) "Materials
with Novel Electronic Properties - MaNEP".

\bibliography{references}
\bibliographystyle{unsrt}

\end{document}